\documentclass[12pt]{iopart}

\usepackage{epsfig}
\usepackage{graphicx}
\usepackage{bm}
\usepackage{iopams}
\usepackage[sort]{cite}

\expandafter\let\csname equation*\endcsname\relax

\expandafter\let\csname endequation*\endcsname\relax

\usepackage{amsmath}
\usepackage{lscape}
\usepackage[usenames,dvipsnames]{color} 
\usepackage[normalem]{ulem}

\renewcommand{\phi}{\varphi}
\newcommand{\be}{\begin{equation}}
\newcommand{\ee}{\end{equation}}
\newcommand{\bea}{\begin{equnaray}}
\newcommand{\eea}{\end{equnaray}}
\newcommand{\ba}{\begin{align}}
\newcommand{\ea}{\end{align}}
\newcommand{\ave}[1]{\left\langle {#1} \right\rangle}

\usepackage{color}

\begin{document}

\title[Classification of mobile- and immobile-molecule timescales in supercooled water]
{Classification of mobile- and immobile-molecule timescales for the Stokes--Einstein and Stokes--Einstein--Debye relations in supercooled water}

\author{Takeshi Kawasaki$^{1}$ and Kang Kim$^{2}$}
\address{$^1$ Department of Physics, Nagoya University, Nagoya 464-8602, Japan}
\address{$^2$ Division of Chemical Engineering, Graduate School of Engineering Science, Osaka University, Toyonaka, Osaka 560-8531, Japan}
\ead{kawasaki@r.phys.nagoya-u.ac.jp and kk@cheng.es.osaka-u.ac.jp}
\date{\today}% It is always \today, today,

\begin{abstract}
Molecular dynamics simulations have been performed on TIP4P/2005
supercooled water to investigate the molecular diffusion and shear
viscosity at various timescales and assess the Stokes--Einstein (SE) and
Stokes--Einstein--Debye (SED) relations.
For this purpose, we calculated various time correlation functions, such
as the mean-squared displacement, stress relaxation function, density
correlation function, hydrogen-bond correlation function, rotational
correlation function of molecular orientation, non-Gaussian parameter,
and four-point correlation function. 
Our study of the SE and SED
relations indicates that the transport coefficients and timescales
obtained using these time correlation functions may be classified into
two distinct classes: those governed by either mobile or immobile
molecules, due to dynamical heterogeneity.
In particular, we show that
the stress relaxation time, hydrogen-bond lifetime, and large-angle
rotational relaxation time are coupled with translational diffusion, and
are characterized by mobile molecules.
In contrast, the structural
$\alpha$-relaxation time, small-angle rotational relaxation time, and
characteristic timescales of four-point correlation functions are
decoupled with translational diffusion, and are governed by immobile
molecules.
This decoupling results in a violation of the SE
relation.
These results indicate that the identification of timescales
that appropriately characterize transport coefficients, such as
translational diffusion constant and shear viscosity, provides a deep
insight into the violation of the SE and SED relations in glass-forming
liquids.

\end{abstract}

\vspace{2pc}
\noindent{\it Keywords}: slow relaxation and glassy dynamics, Molecular
dynamics, Dynamical heterogeneities
%\submitto{\JSTAT}
%\maketitle

\section{Introduction}

The need for a unified description of the structural relaxation
mechanism in
various glass-forming liquids is one of the fundamental problems of
condensed matter physics~\cite{Ediger:1996dz, Ediger:2000ed}.
Scattering and spectroscopic experiments provide data for various timescales, such
as the structural $\alpha$-relaxation time $\tau_\alpha$ of the density
correlation function and relaxation times for
molecular reorientation $\tau_\ell$, obtained using the $\ell$-th order of the Legendre polynomial.
Moreover, transport coefficients such as shear viscosity and diffusion
constant play a crucial role in characterizing the slow dynamics of
glass-forming liquids in the vicinity of the glass transition
temperature~\cite{Angell:1988ia, Debenedetti:2001bh}

The Stokes--Einstein (SE) relation, ${D_\mathrm{t}}^{-1}\propto \eta/T$, and
the Stokes--Einstein--Debye (SED) relation, $\tau_\ell\propto \eta/T$ (or
${D_\mathrm{t}}^{-1}\propto \tau_\ell$), are thought to comprise the key
characteristics necessary to provide the required unified description of
structural relaxation.
Here, $T$, $\eta$, and $D_\mathrm{t}$ denote the temperature, the shear viscosity, and the
translational diffusion constant, respectively.
In fact, these SE and SED relations
typically break down in the case of supercooled states, which is regarded as a hallmark
of the spatially heterogeneous dynamics that are characterized by the
non-Gaussian and the non-exponential nature of various time correlation
functions~\cite{Fujara:1992ib, Cicerone:1996cb, Ediger:2000ed}.
In particular, the violations of SE and SED relations indicate 
decouplings between the molecular diffusion and
the shear viscosity~\cite{Hodgdon:1993ew, Stillinger:1994ko, Tarjus:1995gx, Ngai:2009ge}.
Thus, it is necessary to 
reveal the link between transport coefficients and the
characteristic timescales, and many theoretical studies have been devoted
to this issue~\cite{Hansen:2006vw, Balucani:1995ud}.

Molecular dynamics (MD) simulations of model glass-forming liquids have
been previously conducted in order to investigate the violation of the SE
relation~\cite{Thirumalai:1993cv, Yamamoto:1998jg, Yamamoto:1998gb, Horbach:1999ib,
Bordat:2003cs, Berthier:2004ch, Kumar:2006kr, Kim:2010ii,
Ikeda:2011iq, Shi:2013ji, Sengupta:2013dg,
Kawasaki:2013bg, Kawasaki:2014ky, Henritzi:2015jpa, Saw:2015ii,
Ozawa:2016bk, Schober:2016im, Parmar:2017hl, Puosi:2018cm}.
Whether the temperature dependence of
shear viscosity $\eta$ may be replaced by the $\alpha$-relaxation time
$\tau_\alpha$ is a controversial issue~\cite{Shi:2013ji}.
However, numerical calculations of $\eta$ have demonstrated $\tau_\alpha\propto
\eta/T$ in both fragile and strong glass-formers, indicating that
$D_\mathrm{t}\tau_\alpha$ is a good indicator for $D_\mathrm{t}\eta/T$, even in supercooled
states~\cite{Yamamoto:1998gb, Kawasaki:2014ky}.
The violation of the SED relation in supercooled molecular liquids has
also been examined using MD simulations~\cite{Kammerer:1997ku,
Lombardo:2006jq, Chong:2009ci}.
However,
there has been no direct assessment of the quantity $\tau_\ell T/\eta$
through the variation of $\ell$.
Instead, the temperature dependence of $D_\mathrm{t}\tau_\ell$
with $\ell=2$ has been mainly quantified, which is comparable to the
results of experimental analysis~\cite{Fujara:1992ib, Cicerone:1996cb}.
Furthermore, the connection between $\tau_\alpha$ and $\tau_\ell$ for 
varying the degree $\ell$ 
remains elusive, particularly in supercooled states, and thus particular care
should be taken when discussing the coupling of translational and rotational
molecular motions.

The system of supercooled water has attracted much attention owing to
its capacity for elucidating the 
mechanism underlying SE and SED relations through both experiments
and simulations~\cite{Chen:2006kk, Becker:2006ju, Kumar:2006hx, Kumar:2007hl, Mazza:2007kr,
Xu:2009hq, Banerjee:2009db, Mallamace:2010uj, Jana:2011fj, Qvist:2012gg,
Rozmanov:2012ja, Bove:2013em, Dehaoui:2015ii,
Guillaud:2017bk, Guillaud:2017ey,
Galamba:2017eq, Kawasaki:2017gw, Shi:2018gu, MonterodeHijes:2018ec,
Saito:2018dn, Kawasaki:2018vv}.
Comprehensive numerical calculations of the shear
viscosity enabled the precise evaluation of SE and SED
relations~\cite{Kawasaki:2017gw, Kawasaki:2018vv}.
In particular, it was demonstrated that
the shear viscosity $\eta$ can be represented by the approximation 
$\eta \approx G_\mathrm{p}\tau_\eta\Gamma(1/\beta_\eta)/\beta_\eta$, with
the Gamma function $\Gamma(x)$, according to the approximation,
$G_\eta(t)\approx G_\mathrm{p}\exp\{-(t/\tau_\eta)^{\beta_\eta}\}$, for
the long time behavior of $G_\eta(t)$~\cite{Kawasaki:2017gw}.
Here, $G_\mathrm{p}$, $\tau_\eta$, and $\beta_\eta$ denote the plateau
modulus, the stress relaxation time, and  the degree of
non-exponentiality 
of the stress correlation function $G_\eta(t)$, respectively.
In contrast, the translational diffusion constant $D_\mathrm{t}$ is
governed by hydrogen-bond (H-bond) breakage processes,
leading to the proportional relationship of $D_\mathrm{t}^{-1} \propto
\tau_\mathrm{HB}$ with the H-bond lifetime, $\tau_\mathrm{HB}$~\cite{Kawasaki:2017gw}.
An analogous relationship has been demonstrated in both
fragile supercooled liquids~\cite{Kawasaki:2013bg} and
silica-like strong supercooled liquids~\cite{Kawasaki:2014ky}, where
the bond-breakage method was used to characterize changes in local
connectivity of molecules~\cite{Yamamoto:1997gu, Yamamoto:1998jg,
Shiba:2012hm, Shiba:2016bi, Shiba:2018ip}.
Furthermore, we previously conducted a comprehensive investigation into the SED relations,
$\tau_\ell T/\eta$ and $D_\mathrm{t}\tau_\ell$, for various $\tau_\ell$~\cite{Kawasaki:2018vv}.
It was demonstrated that these SED relations shows a strong dependence on the
degree $\ell$: the higher-order $\tau_\ell$ values exhibit a
temperature dependence similar to that of $\eta/T$, whereas the
lowest-order $\tau_\ell$ values are coupled with $D_\mathrm{t}$.

In this study, we examine the roles of characteristic timescales, including
the $\alpha$-relaxation time $\tau_\alpha$, H-bond lifetime
$\tau_\mathrm{HB}$, rotational relaxation times $\tau_\ell$ ($\ell=1$ and
6), stress relaxation time $\tau_\eta$, and the timescales of non-Gaussian
parameters and four-point dynamic correlations, in the SE and SED relations.
In particular, the aim is to classify these timescales
into two classes: those are coupled or decoupled with $D_\mathrm{t}$ (or $\eta/T$).
We also discuss this classification in terms of the mobile/immobile
contributions of the dynamic heterogeneities in supercooled water.
The rest structure of the paper is organized as follows:
In Sec.~\ref{methods}, we describe the MD simulations of supercooled water using the
TIP4P/2005 model.
In Sec.~\ref{results}, we describe the numerical calculations of various timescales
and discuss their temperature dependence, compared with those of
$D_\mathrm{t}$ and $\eta/T$.
In Sec.~\ref{summary}, we summarize our conclusions.

\section {MD Simulations}
\label{methods}

We performed MD simulations
of liquid water using the Large-scale Atomic/Molecular Massively
Parallel Simulator (LAMMPS)~\cite{Plimpton:1995wl}, and used the
TIP4P/2005 water model~\cite{Abascal:2005ka}. 
There are various studies in the literature of MD
simulations used to investigate
various properties in supercooled states of this
model~\cite{Abascal:2010dw, Sumi:2013fy, Overduin:2013cu,
DeMarzio:2016hl, Hamm:2016hf, Singh:2016bu, Gonzalez:2016gr,
Handle:2018cn, MonterodeHijes:2018ec, Saito:2018dn}.
In the present study, 
in order to obtain equilibrated initial configurations, we performed
simulations with the $NVT$ ensembles for $N = 1,000$ water molecules at
various temperatures ($T = 300, 280, 260, 250, 240, 230, 220, 210, 200$,
and 190 K) at a fixed mass density of $\rho = 1$ g cm$^{-3}$. 
The corresponding linear dimension was $L = 31.04$ \AA. 
The $NVE$ ensemble simulations were conducted after the
equilibration, remaining at each temperature for a sufficiently
long-time periods, from
which the various time correlation functions were calculated.
Periodic boundary conditions were utilized in all the simulations, with
a simulation time step of 1 fs.
%01
\begin{figure*}[t]
\centering
\includegraphics[width=0.8\columnwidth]{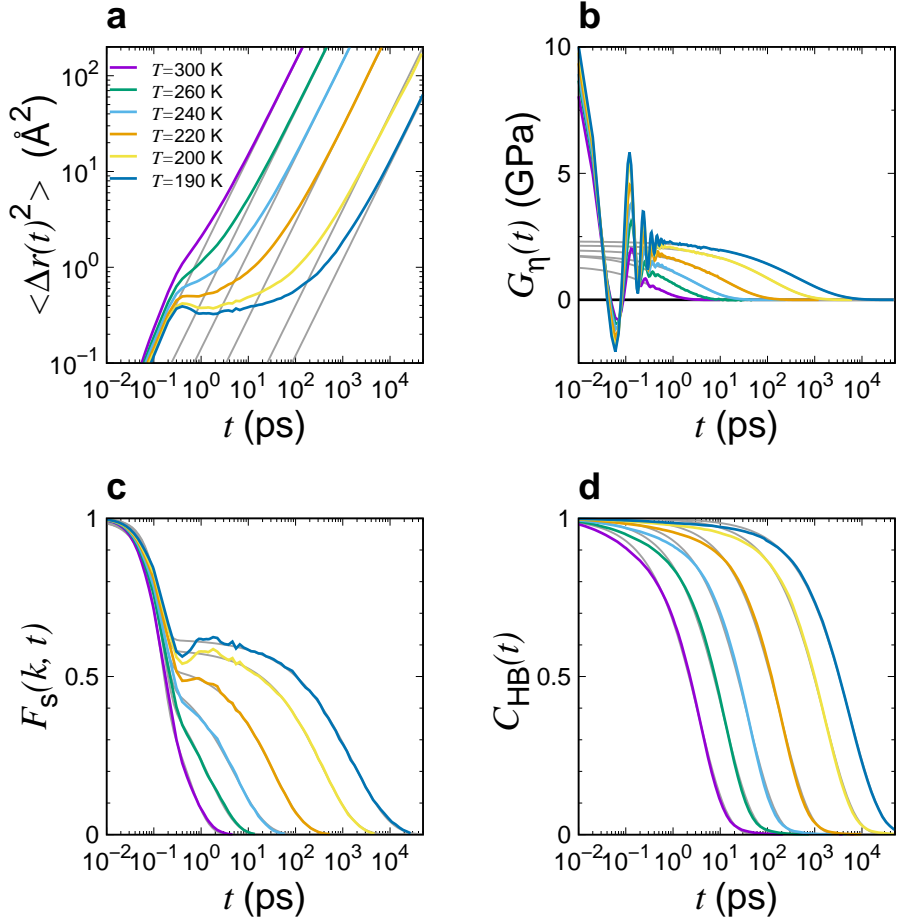}
\caption{
(a) Mean-squared displacement $\ave{\Delta r
 (t)^2}$.
Gray solid lines represent $6D_\mathrm{t}t$ obtained by the fitting of the
 long time region in $\ave{\Delta r (t)^2}$.
(b) The Stress correlation function $G_{\eta}(t)$. 
Stress relaxation time $\tau_{\eta}$ is determined by the fitting
 $G_{\eta}(t)$ with the KWW function,
 $G_\mathrm{p}\exp{\{-(t/\tau_{\eta})^{\beta_\eta}\}}$. 
Gray solid lines represent the fitting results. 
(c) The incoherent intermediate scattering function
 $F_\mathrm{s}(k,t)$.
The structural $\alpha$-relaxation time $\tau_{\alpha}$ is determined by the
 fitting $F_\mathrm{s}(k,t)$ with the KWW function
$(1-f_\mathrm{c}) \exp[-(t/\tau_\mathrm{s})^2] + f_\mathrm{c}
 \exp[-(t/\tau_\alpha)^{\beta_\alpha}]$. 
Gray solid lines represent the fitting results.
(d) H-bond correlation function $C_\mathrm{HB}(t)$.
The H-bond lifetime $\tau_\mathrm{HB}$ is determined by fitting $C_\mathrm{HB}(t)$ 
with $\exp[-(t/\tau_\mathrm{HB})^{\beta_\mathrm{HB}}]$. 
Gray solid lines represent the fitting results.}
\label{fig1} 
\end{figure*}

%02
\begin{figure*}[t]
\centering
\includegraphics[width=0.8\columnwidth]{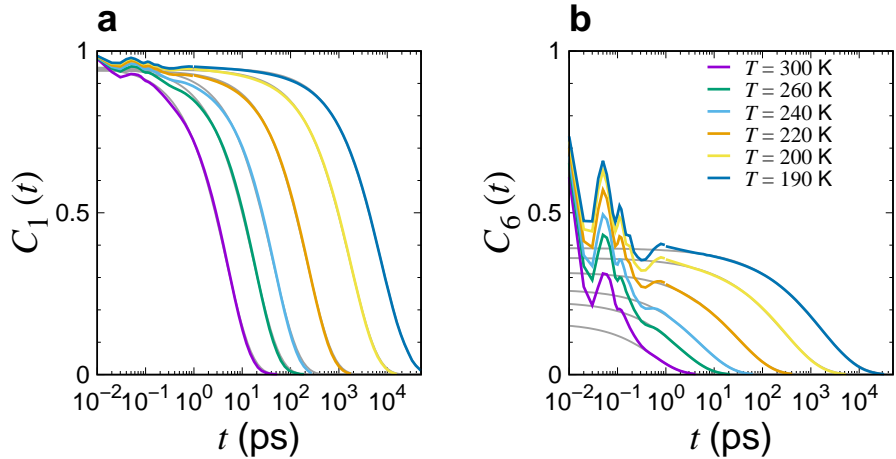}
\caption{
Rotational correlation functions $C_{\ell}(t)$ with $\ell = 1$ (a) and
 $\ell = 6$ (b).
The rotational relaxation time $\tau_\ell$ is determined by fitting
 $C_\ell(t)$ with the KWW function $A_{\ell}\exp{\{-(t/\tau_\ell)^{\beta_\ell}\}}$. 
}
\label{fig2} 
\end{figure*}

%03
\begin{figure*}[t]
\centering
\includegraphics[width=0.8\columnwidth]{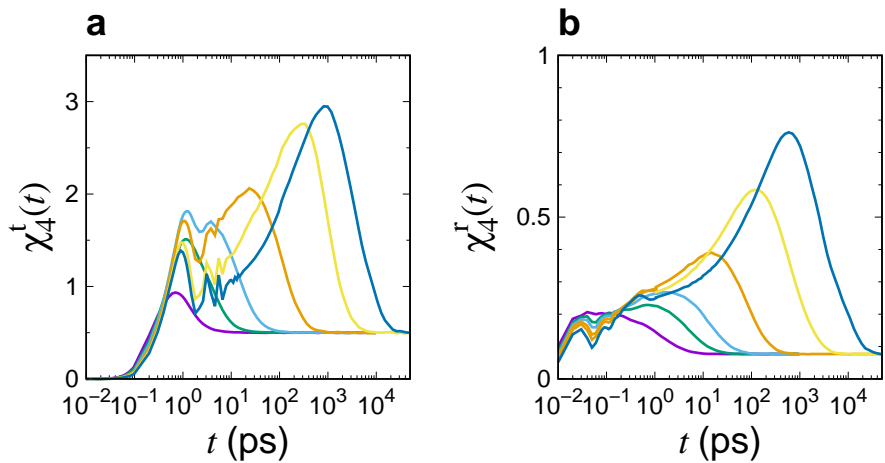}
\caption{
Four-point correlation functions of translational (a) and rotational (b) motion.
The peak times of $\chi_4^\mathrm{t}(t)$ and $\chi_4^\mathrm{r}(t)$ are
 denoted by $\tau_\mathrm{t}$ and $\chi_4^\mathrm{r}(t)$, respectively.
}
\label{fig3} 
\end{figure*}

%04
\begin{figure*}[t]
\centering
\includegraphics[width=0.8\columnwidth]{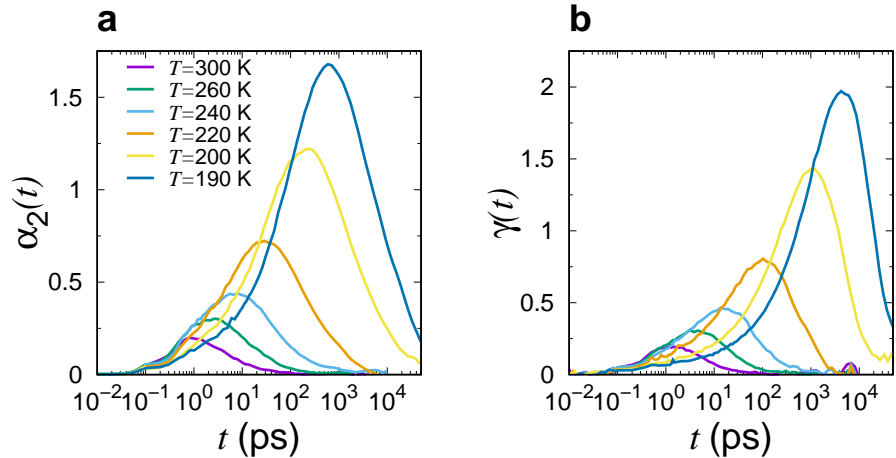}
\caption{
Non-Gaussian parameter $\alpha_2(t)$ (a) and new non-Gaussian
 parameter $\gamma(t)$ (b).
The peak times of $\alpha_2(t)$ and $\gamma(t)$ are denoted by
 $\tau_\mathrm{ngp}$ and $\tau_\mathrm{nngp}$, respectively.
}
\label{fig4} 
\end{figure*}

%05
\begin{figure*}[t]
\centering
\includegraphics[width=0.95\columnwidth]{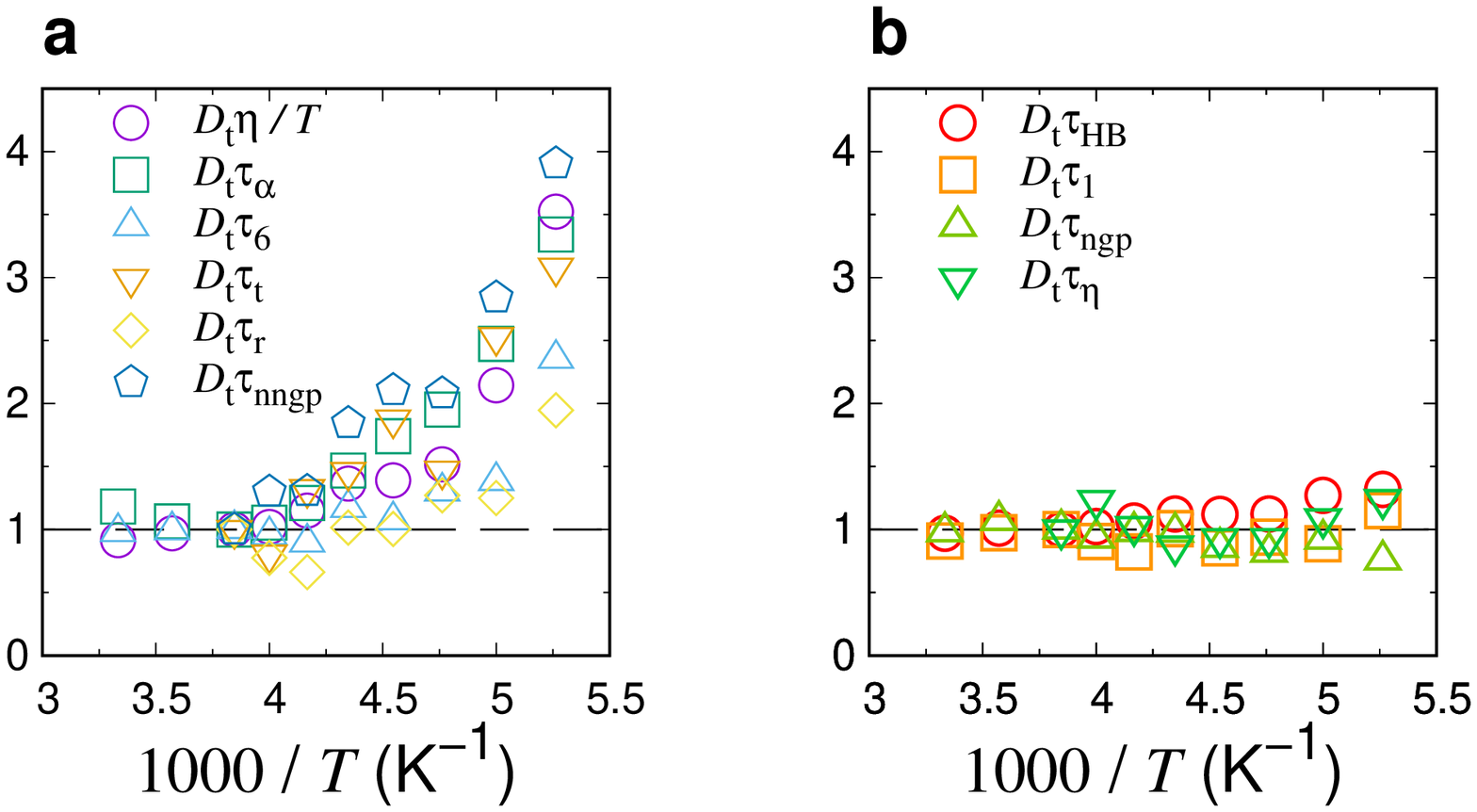}
\caption{
(a) Inverse temperature $1000/T$ dependence of the scaled SE ratios,
 obtained using
 shear viscosity divided by temperature $\eta/T$, 
 structural $\alpha$-relaxation time $\tau_{\alpha}$, rotational relaxation
 time $\tau_6$, peak time of the translational 
 four-point correlation function $\tau_\mathrm{t}$, peak
 time of the rotational four-point correlation function $\tau_\mathrm{r}$, and
 peak time of NNGP $\tau_\mathrm{nngp}$. 
All quantities are scaled by their values at 260 K.
(b)  Inverse temperature $1000/T$ dependence of the scaled SE ratios obtained
 using H-bond life time $\tau_\mathrm{HB}$, rotational
 relaxation time $\tau_1$, peak time of NGP $\tau_\mathrm{ngp}$, and 
stress relaxation time $\tau_{\eta}$. 
All quantities are scaled by their values at 260 K.
}
\label{fig5} 
\end{figure*}

\section{Results and discussion}
\label{results}

\subsection{Mean-squared displacement and translational diffusion constant}

As an important component of the SE relation,
we quantified the translational diffusion constant $D_\mathrm{t}$. 
To this end, we calculated the mean squared displacement,
\be
\langle \Delta r(t)^2\rangle =
\left\langle\frac{1}{N}\sum_{j=1}^N|\mathbf{r}_j(t)-\mathbf{r}_j(0)|^2\right\rangle,
\ee
where $\mathbf{r}_j(t)$ represents the position of $j$-th O atom at time $t$.
The bracket $\ave{\cdots}$ indicates an average over the initial time 0. 
In Fig.~\ref{fig1}(a), $\ave{\Delta r (t)^2}$ is plotted for various temperatures.
For each temperature, the ballistic
behavior, $\ave{\Delta r (t)^2}\propto t^2$, is observed in the short time region.
The diffusive behavior, $\ave{\Delta r(t)^2}\propto t$, eventually
develops in the long time region.
The plateau that is observed in the intermediate region is pronounced upon
supercooling, indicating cage effects related to the presence of neighboring molecules. 
The translational diffusion constant $D_\mathrm{t}$ is determined from the
long time behavior of $\langle \Delta r(t)^2\rangle$ using the Einstein relation, 
$D_\mathrm{t}=\lim_{t\to\infty}\langle \Delta r(t)^2\rangle / 6t$.
The solid lines in Fig.~\ref{fig1} (a) are the results of the fitting
with 
$\ave{\Delta r (t)^2}=6D_\mathrm{t}t$. 

\subsection{Stress correlation function and shear viscosity}

In order to obtain the shear viscosity $\eta$ as another essential
component of the SE relation, the autocorrelation function
of the off-diagonal stress tensor was calculated, which is given by
\be
G_{\alpha\beta}(t) = \frac{V}{k_\mathrm{B}T} \langle \sigma_{\alpha\beta}(t)
 \sigma_{\alpha\beta}(0)\rangle,
\ee
where $V$ is the volume of the system and $\sigma_{\alpha\beta}$
represents the $\alpha\beta(=x, y, z)$ components of the off-diagonal
stress tensor. 
The average stress correlation function is then defined as
$G_\eta(t) = (G_{xy}(t)+G_{xz}(t)+G_{yz}(t))/3$. 
$G_\eta(t)$ is plotted in Fig.~\ref{fig1}~(b), for various temperatures.
The instantaneous time correlation $G_\eta(0)$ corresponds to the
instantaneous or affine shear modulus $G_{\infty}$. 
In the short time region, $G_\eta(t)$ shows large fluctuations, which are 
attributed to vibrational motions observed in network forming liquids
such as silica glass~\cite{Kawasaki:2014ky}.
The $G_\eta(t)$ plateaus in the intermediate time region,
particularly at lower temperatures, which is the so-called the plateau modulus.
$G_\eta(t)$ finally decays to zero at longer timescales.
The shear viscosity $\eta$ is determined from the integral of $G_\eta(t)$ as
\be
\eta = \int_0^\infty G_\eta (t) \mathrm{d}t,
\ee
using the Green--Kubo formula.
Furthermore, we examine the stress relaxation time $\tau_{\eta}$ of $G_\eta(t)$.
The long time behavior of $G_\eta(t)$ is fitted using the
Kohlrausch--Williams--Watts (KWW) stretched-exponential function,
$G_\mathrm{p}\exp\{-(t/\tau_\eta)^{\beta_\eta}\}$.
Here, $G_\mathrm{p}$ and $\tau_\eta$ denote
the plateau modulus and the stress relaxation time, respectively.
The exponent $\beta_\eta(< 1)$ relates to the degree of non-exponentiality of
$G_\eta(t)$. 
The gray solid lines in Fig.~\ref{fig1}(b) indicate those fitting results.

The SE ratio $D_\mathrm{t}\eta /T$ as a function of the
inverse of the temperature is displayed in
Fig.~\ref{fig5}(a).
This plot demonstrates the violation of the SE relation for 
temperatures below $240$ K.

\subsection{Incoherent intermediate scattering function and $\alpha$-relaxation time}

As an alternative to $\eta$, the structural $\alpha$-relaxation
time $\tau_{\alpha}$ has been frequently used in discussions of the
violation of the SE relation, which is obtained from
the incoherent intermediate scattering function given by
\be
F_\mathrm{s}(k, t)=\left\langle\frac{1}{N}
\sum_{j=1}^N\exp[i\mathbf{k}\cdot(\mathbf{r}_j(t)-\mathbf{r}_j(0))]\right\rangle.
\ee
Here, $\mathbf{r}_j(t)$ is the position vector of the O atom of the water molecule $j$ at time $t$. 
The wave number $k=|\mathbf{k}|$ was set to $k=3.0$ \AA ${}^{-1}$, which corresponds to 
the first peak position of the static structure factors $S(k)$ of the O atom. 
$F_\mathrm{s}(k, t)$ is plotted for various temperatures in Fig.~\ref{fig1}~(c).
A two-step relaxation in $F_s(k, t)$ is observed upon supercooling.
The $\alpha$-relaxation time $\tau_\alpha$ is determined by the fitting
$F_\mathrm{s}(k, t)$ with the KWW function, {\it i.e.,} $(1-f_\mathrm{c})
\exp[-(t/\tau_\mathrm{s})^2] + f_\mathrm{c}
\exp[-(t/\tau_\alpha)^{\beta_\alpha}]$, where $f_\mathrm{c}$,
$\tau_\mathrm{s}$, $\tau_\alpha$, and $\beta_\alpha$ are the fitting
parameters~\cite{Sengupta:2013dg}. 
The exponent $\beta_{\alpha}$ is the degree of
non-exponentiality of $F_\mathrm{s}(k, t)$. 

The quantity $D_\mathrm{t}\tau_{\alpha}$ is plotted  in Fig.~\ref{fig5}(a) for various temperatures.
These results show that the violation of the SE relation is explained
in terms of 
$D_\mathrm{t}\tau_{\alpha}$, suggesting the proportional relationship, $\tau_\alpha \propto \eta/T$.

\subsection {Hydrogen-bond breakage and its lifetime}

Next, we focus on the H-bond life time $\tau_\mathrm{HB}$, which is thought
to be a type of structural relaxation of molecules in the case of liquid
water.
However, it will be demonstrated 
that $\tau_\mathrm{HB}$ behaves differently from $\eta/T$ and $\tau_{\alpha}$.
The dynamics of H-bond network rearrangement was investigated using the correlation function,
\be
C_\mathrm{HB}(t) = \left\langle \sum_{i,
j}^N h_{ij}(0)h_{ij}(t)\right\rangle \bigg/ \left\langle \sum_{i, j}^N h_{ij}(0)h_{ij}(0)\right\rangle, 
\ee
where $h_{ij}(t)$ denotes the H-bond indicator between $i$-th and $j$-th molecules at a
time $t$~\cite{Rapaport:1983ib, Saito:1995gz, Luzar:1996gw,
Luzar:1996gx, Luzar:2000gv}.
In this study, the H-bond is defined by the 
r-definition, where only the intermolecular O-H
distance $r_\mathrm{OH}$ is investigated~\cite{Kumar:2007bs}. 
$h_{ij}(t)=1$
if the $r_\mathrm{OH}$ between the $i$-th and $j$-th molecules is less than $2.4$ \AA, corresponding to
the first minimum of the radial distribution function
$g_\mathrm{OH}(r)$, and $h_{ij}(t)=0$ otherwise.
The time dependence of $C_\mathrm{HB}(t)$ is given in
Fig.~\ref{fig1}(d), where
the decay time of
$C_\mathrm{HB}(t)$ increases with decreasing the temperature.
However, 
they do not show the two-step relaxation for all temperatures, contrary
to the observations of $F_\mathrm{s}(k,t)$.
The H-bond lifetime
$\tau_\mathrm{HB}$ is determined by the fitting $C_\mathrm{HB}(t)$ with the KWW
functions, $\exp\{-(t/\tau_\mathrm{HB})^{\beta_\mathrm{HB}}\}$, where the
exponent $\beta_\mathrm{HB}$ is the degree of non-exponentiality of
$C_\mathrm{HB}(t)$.
The gray solid lines in Fig.~\ref{fig1}~(d) indicate the results of the KWW fittings.
Note that the H-bond correlation function $C_\mathrm{HB}$ has a form
that is essentially identical to the bond-breakage method, which
has been applied to various supercooled liquids~\cite{Yamamoto:1997gu, Yamamoto:1998jg,
Shiba:2012hm, Kawasaki:2013bg, Kawasaki:2014ky,Shiba:2016bi, Shiba:2018ip}.
These previous studies have demonstrated that the
bond-breakage method is a remarkable method for characterizing ``genuine''
configurational rearrangements in supercooled states.

Figure~\ref{fig5}(b) provides the temperature dependence of
$D_\mathrm{t}\tau_\mathrm{HB}$ for various temperatures.
Remarkably, these plots reveal that $\tau_\mathrm{HB}$ is coupled with
$D_\mathrm{t}$, which differs from $\tau_\alpha$.
This result can be thought of as the preservation of the SE relation~\cite{Kawasaki:2017gw}.
Furthermore, this preservation of the SE relation is 
interpreted in terms of the observation that 
$D_\mathrm{t}$ and $\tau_\mathrm{HB}$ are governed by  
molecules with jump/fast motions~\cite{Kawasaki:2013bg, Kawasaki:2014ky,
Kawasaki:2017gw}.
In contrast,
we will show that $\eta$ and $\tau_{\alpha}$ are governed by immobile
molecules with dynamic heterogeneities.

\subsection{Rotational relaxation times}

Molecular reorientations are often characterized by the rotational
correlation function,
\be
C_{\ell}(t) = \frac{1}{N}\sum_{j=1}^N \ave{P_{\ell}[\mathbf{e}_j(t) \cdot \mathbf{e}_j(0)]}, 
\ee
where $\mathbf{e}_j(t)$ is the normalized
polarization vector of molecule $j$ and 
$P_{\ell}[x]$ is the $\ell$-th order Legendre polynomial as a
function of $x$. 
In the present study, we examine the $\ell$-th ($\ell=1$ and $6$) order
rotational relaxation function $C_{\ell}(t)$.
Plots of $C_{\ell}(t)$ are provided in Fig.~\ref{fig2}(a) and (b), with $\ell=1$ and
$6$, respectively.
The gray solid lines represent the fitting results of the KWW function,
$A_{\ell}\exp{\{-(t/\tau_{\ell})^{\beta_{\ell}}\}}$.
The rotational relaxation time $\tau_1$ and $\tau_6$ were quantified as
a result of this fitting procedure.

The SED ratio $D_\mathrm{t} \tau_6$ is given in Fig.~\ref{fig5}(a) as
a function of the inverse of the temperature.
This result shows the violation of the SE relation.
However, another ratio
$D_\mathrm{t} \tau_1$ indicates the preservation of the SED relation, as shown
in Fig.~\ref{fig5}(b).
This conflicting result can be understood as follows:
$\tau_1$ is characterized by the molecular reorientation related to 
large rotational movements, which are mostly large
translational jump motions. 
Therefore, $\tau_1$ is coupled with $D_\mathrm{t}$.
On the contrary, the small angular
rotational movements characterizing $\tau_6$ are governed by
``immobile'' molecules, which is decoupled with $D_\mathrm{t}$,
particularly at lower temperatures.

\subsection{Characterizations of dynamic heterogeneities -- four-point correlation functions}

In order to elucidate the degree of dynamic heterogeneities and the
related 
characteristic timescales in supercooled water, 
we examine the
four-points correlation functions for translational and rotational
motions~\cite{Toninelli:2005ci}. 
The four-point correlation function
$\chi_4^\mathrm{t}(t)$ for translational motion is defined by the
variance of the intermediate scattering function $F_s(k, t)$ as,
\be
\chi_4^\mathrm{t}(t) = N\left[\langle \hat F_s(k, t)^2\rangle -\langle
 \hat F_s(k, t)\rangle^2 \right], 
\ee
where $\hat F_s(t)=(1/N)\sum_{j=1}^N\cos[{\bf k}\cdot \Delta {\bf r}_j(t)]$.
The wave number $k=|{\bf k}|$ was set to $k=$3.0 \AA${}^{-1}$.
The rotational four-point correlation
function $\chi_4^\mathrm{r}(t)$ can be analogously defined as
\be
\chi_4^\mathrm{r}(t) = N\left[\langle \hat C_\ell(t)^2\rangle -\langle
\hat C_\ell(t)\rangle^2\right],
\ee
where $\hat C_\ell(t) = (1/N)\sum_{j=1}^N P_\ell[\mathbf{e}_j(t)\cdot \mathbf{e}_j(0)]$. 
The order of the Legendre polynomial was set to $\ell=6$.
Figure~\ref{fig3} shows the temperature dependence of the four-point
correlation functions, $\chi_4^\mathrm{t}(t)$ and $\chi_4^\mathrm{r}(t)$. 
The peak times of $\chi_4^\mathrm{t}(t)$ and $\chi_4^\mathrm{r}(t)$ are
quantified and denoted by $\tau_\mathrm{t}$ and $\tau_\mathrm{r}$, respectively.
We find that upon supercooling the peak heights increase, together with
an increase in their peak times $\tau_\mathrm{t}$  and  $\tau_\mathrm{r}$ 
are also increased.

Figure~\ref{fig5}(a) shows a plot of 
$D_\mathrm{t}\tau_{t}$  and  $D_\mathrm{t}\tau_\mathrm{r}$
as a function of the inverse of the temperature.
Both quantities exhibit a temperature dependence similar to that
observed in the case of the violation of the SE relation.
This indicates that the characteristic timescales of translational and
rotational dynamic heterogeneities are coupled with $\eta/T$ and $\tau_{\alpha}$.

\subsection{Characterizations of dynamic heterogeneities -- non Gaussian parameters}

Finally, we elucidate the role of ``mobile'' and ``immobile'' molecules
with dynamical heterogeneities on the SE and SED relations.
First, we investigate the
non-Gaussian parameter (NGP) that mainly characterizes the displacements of ``mobile''
molecules~\cite{Rahman:1964ht, Kob:1997hl}.
The equation is given by 
\be
\alpha_2(t) =  \frac{3}{5}\frac{\langle \Delta r(t)^4\rangle}{\langle
 \Delta r(t)^2 \rangle^2}-1.
\ee
Second, to emphasize the effect of ``immobile'' molecules, another type
of non-Gaussian parameter is defined by
\be
\gamma(t) =  \frac{1}{3}{\langle \Delta r(t)^2\rangle}\left\langle
 \frac{1}{\Delta r(t)^2} \right\rangle-1, 
\ee
which is referred to as the new non-Gaussian parameter (NNGP)~\cite{Flenner:2005es}.
Figure~\ref{fig4} shows the time evolutions of NGP $\alpha_2(t)$ and NNGP
$\gamma(t)$ at various temperatures.
The peak heights of $\alpha_2(t)$ and $\gamma(t)$ increase with
decreasing the temperature.
This means that the dynamics are more heterogeneous upon supercooling. 
Their peak times are denoted by $\tau_\mathrm{ngp}$ and
$\tau_\mathrm{nngp}$, respectively.
The corresponding temperature dependences $D_\mathrm{t}\tau_\mathrm{ngp}$ and
$D_\mathrm{t}\tau_\mathrm{nngp}$ are given in Fig.~\ref{fig5}(a) and
Fig.~\ref{fig5}(b), respectively.
As shown in Fig.~\ref{fig5}(a),
the timescale of the ``immobile'' molecules, $\tau_\mathrm{nngp}$, follows the
temperature dependence observed for the violation of the SE relation,
$D_\mathrm{t}\eta/T$ and $D_\mathrm{t}\tau_\alpha$.
It is eventually concluded that the timescales, $\tau_\alpha$, $\tau_6$,
$\tau_\mathrm{t}$, and $\tau_\mathrm{r}$ are dominated by ``immobile''
molecules of dynamic heterogeneities.
By contrast,
Fig.~\ref{fig5}(b) demonstrates that $\tau_\mathrm{ngp}$ follows the preservation of
the SE relation.
In this case, the timescales that are coupled with $D_\mathrm{t}$, $\tau_{\eta}$, $\tau_\mathrm{HB}$, and
$\tau_{1}$, are dominated by ``mobile'' molecules with dynamic heterogeneities.

\section{Conclusion}
\label{summary}

In this study, we performed MD simulations on supercooled liquid water using the TIP4P/2005 model.
We determined the temperature dependence of transport coefficients such
as the translational diffusion constant
$D_\mathrm{t}$ and shear viscosity $\eta$, together with 
the timescales characterizing the dramatic slowing down of glassy dynamics.
The SE relation, $D_\mathrm{t}\eta/T$, was also thoroughly assessed.
The temperature dependence of two transport coefficients are completely
decoupled with decreasing the temperature, suggesting the violation of the
SE relation in supercooled water.
Furthermore, the SED relations $\tau_\ell T/\eta$ and
$D_\mathrm{t}\tau_\ell$ were investigated.
We determined 
that these SED relations exhibit a strong dependence on the order $\ell$ of the
Legendre polynomial, \textit{i.e.}, the examined angle of molecular reorientations.

These assessments of the SE and SED relations enabled the classification of
various characteristic timescales into just two classes;
dominated by ``mobile'' and ``immobile'' molecules of dynamical
heterogeneities, depending on the degree of coupling with 
the translational diffusion constant $D_\mathrm{t}$.
Moreover, $D_\mathrm{t}$ is coupled with the H-bond lifetime
$\tau_\mathrm{HB}$ that is governed by mobile molecules that exhibit
large molecular displacement amplitudes.
These coupling dynamics indicate the preservation of the SE relation, 
which was rationalized by the peak time of NGP, $\tau_\mathrm{ngp}$,
which characterize the mobile molecules of dynamic heterogeneities. 
In addition, 
the stress relaxation time $\tau_{\eta}$ and 
the rotational relaxation time $\tau_1$ that is accompanied by a large angle amplitude,
show the temperature dependence similar to 
that of $D_\mathrm{t}$, which is also governed by mobile molecules.

In contrast, 
$\eta/T$ is proportional to 
the structural $\alpha$-relaxation time $\tau_{\alpha}$
that is quantified by the density correlation function.
This implies the
violation of the SE relation (the decoupling between $\eta/T$ and
$D_\mathrm{t}$) that are related to 
immobile molecules of dynamic heterogeneities.
In fact, the peak time of NNGP, $\tau_\mathrm{nngp}$ that characterize
the contribution of immobile molecules
shows the temperature dependence similar to that of the violation of the SE relation.
As discussed, the shear viscosity $\eta$ is represented by 
$\eta \approx G_\mathrm{p}\tau_\eta\Gamma(1/\beta_\eta)/\beta_\eta$
using the Green--Kubo formula~\cite{Kawasaki:2017gw}.
The violation of the SE relation is thus expressed by $D_\mathrm{t}\eta /T \propto G_\mathrm{p}
\Gamma(1/\beta_\eta)/(T\beta_\eta)$, using the observed relationships,
${D_\mathrm{t}}^{-1} \propto \tau_\mathrm{HB}$ and ${D_\mathrm{t}}^{-1} \propto \tau_\eta$.
Furthermore, we show that the 
rotational relaxation time for a small-angle amplitude, $\tau_6$, and the peak times of
both translational and the peak times of rotational four-point correlation
functions, $\tau_\mathrm{t}$ and $\tau_\mathrm{r}$, are 
also governed by immobile molecules.
These are decoupled with $D_\mathrm{t}$, but instead exhibit the 
temperature dependence similar to $\eta/T$.

The identification of timescales
that appropriately characterize the transport coefficients,
$D_\mathrm{t}$ and $\eta$, paves the way to a deeper understanding of
the violation of SE relation that is generally observed in various glass-forming liquids.
In particular, the violation of the SE relation
is directly relevance to the decoupling between the $\alpha$-relaxation time
$\tau_\alpha$ and both the stress relaxation time $\tau_\eta$ and the H-bond
lifetime $\tau_\mathrm{HB}$ in supercooled water.
Further investigations should be undertaken in order to clarify this issue and
obtain a unified description of structural relaxation in glass-forming liquids.

\section*{Acknowledgments}
We thank K. Miyazaki, N. Matubayasi, T. Nakamura, and H. Shiba for 
valuable discussions. 
This work was partly supported by JSPS KAKENHI Grant Numbers JP15H06263,
JP16H06018, and JP18H01188.
The numerical calculations were performed at the Research Center of
Computational Science, Okazaki, Japan.

\section*{References}

\bibliographystyle{iopart-num} 
%\bibliography{kkims2018}

\begin{thebibliography}{10}
\expandafter\ifx\csname url\endcsname\relax
  \def\url#1{{\tt #1}}\fi
\expandafter\ifx\csname urlprefix\endcsname\relax\def\urlprefix{URL }\fi
\providecommand{\eprint}[2][]{\url{#2}}

\bibitem{Ediger:1996dz}
Ediger M~D, Angell C~A and Nagel S~R {\em {Supercooled Liquids and Glasses}\/}
  1996 {\em J. Phys. Chem.\/} {\bf 100}, 13200--13212

\bibitem{Ediger:2000ed}
Ediger M~D {\em {Spatially heterogeneous dynamics in supercooled liquids.}\/}
  2000 {\em Annu. Rev. Phys. Chem.\/} {\bf 51}, 99--128

\bibitem{Angell:1988ia}
Angell C {\em {Perspective on the glass transition}\/} 1988 {\em J. Phys.
  Chemi. Solids\/} {\bf 49}, 863--871

\bibitem{Debenedetti:2001bh}
Debenedetti P~G and Stillinger F~H {\em {Supercooled liquids and the glass
  transition}\/} 2001 {\em Nature\/} {\bf 410}, 259--267

\bibitem{Fujara:1992ib}
Fujara F, Geil B, Sillescu H and Fleischer G {\em {Translational and rotational
  diffusion in supercooled orthoterphenyl close to the glass transition}\/}
  1992 {\em Z. Phys. B\/} {\bf 88}, 195

\bibitem{Cicerone:1996cb}
Cicerone M~T and Ediger M~D {\em {Enhanced translation of probe molecules in
  supercooled o-terphenyl: Signature of spatially heterogeneous dynamics?}\/}
  1996 {\em J. Chem. Phys.\/} {\bf 104}, 7210

\bibitem{Hodgdon:1993ew}
Hodgdon J and Stillinger F {\em {Stokes-Einstein violation in glass-forming
  liquids}\/} 1993 {\em Phys. Rev. E\/} {\bf 48}, 207--213

\bibitem{Stillinger:1994ko}
Stillinger F and Hodgdon J {\em {Translation-rotation paradox for diffusion in
  fragile glass-forming liquids}\/} 1994 {\em Phys. Rev. E\/} {\bf 50},
  2064--2068

\bibitem{Tarjus:1995gx}
Tarjus G and Kivelson D {\em {Breakdown of the Stokes--Einstein relation in
  supercooled liquids}\/} 1995 {\em J. Chem. Phys.\/} {\bf 103}, 3071--3073

\bibitem{Ngai:2009ge}
Ngai K~L {\em {Breakdown of Debye-Stokes-Einstein and Stokes-Einstein relations
  in glass-forming liquids: An explanation from the coupling model}\/} 2009
  {\em Philos. Mag. B\/} {\bf 79}, 1783--1797

\bibitem{Hansen:2006vw}
Hansen J~P and McDonald I~R 2006 {\em {Theory of Simple Liquids, Third
  Edition}\/} (London: Academic Press)

\bibitem{Balucani:1995ud}
Balucani U and Zoppi M 1995 {\em {Dynamics of the Liquid State}\/} Oxford
  series on neutron scattering in condensed matter, 10 (USA: Oxford University
  Press)

\bibitem{Thirumalai:1993cv}
Thirumalai D and Mountain R~D {\em {Activated dynamics, loss of ergodicity, and
  transport in supercooled liquids}\/} 1993 {\em Phys. Rev. E\/} {\bf 47},
  479--489

\bibitem{Yamamoto:1998jg}
Yamamoto R and Onuki A {\em {Dynamics of highly supercooled liquids:
  Heterogeneity, rheology, and diffusion}\/} 1998 {\em Phys. Rev. E\/} {\bf
  58}, 3515--3529

\bibitem{Yamamoto:1998gb}
Yamamoto R and Onuki A {\em {Heterogeneous diffusion in highly supercooled
  liquids}\/} 1998 {\em Phys. Rev. Lett.\/} {\bf 81}, 4915--4918

\bibitem{Horbach:1999ib}
Horbach J and Kob W {\em {Static and dynamic properties of a viscous silica
  melt}\/} 1999 {\em Phys. Rev. B\/} {\bf 60}, 3169--3181

\bibitem{Bordat:2003cs}
Bordat P, Affouard F, Descamps M and M{\"u}ller-Plathe F {\em {The breakdown of
  the Stokes Einstein relation in supercooled binary liquids}\/} 2003 {\em J.
  Phys.: Condens. Matter\/} {\bf 15}, 5397--5407

\bibitem{Berthier:2004ch}
Berthier L {\em {Time and length scales in supercooled liquids}\/} 2004 {\em
  Phys. Rev. E\/} {\bf 69}, 020201(R)

\bibitem{Kumar:2006kr}
Kumar S~K, Szamel G and Douglas J~F {\em {Nature of the breakdown in the
  Stokes-Einstein relationship in a hard sphere fluid}\/} 2006 {\em J. Chem.
  Phys.\/} {\bf 124}, 214501

\bibitem{Kim:2010ii}
Kim K and Saito S {\em {Role of the Lifetime of Dynamical Heterogeneity in the
  Frequency-Dependent Stokes--Einstein Relation of Supercooled Liquids}\/} 2010
  {\em J. Phys. Soc. Jpn.\/} {\bf 79}, 093601

\bibitem{Ikeda:2011iq}
Ikeda A and Miyazaki K {\em {Glass transition of the monodisperse Gaussian core
  model.}\/} 2011 {\em Phys. Rev. Lett.\/} {\bf 106}, 015701

\bibitem{Shi:2013ji}
Shi Z, Debenedetti P~G and Stillinger F~H {\em {Relaxation processes in
  liquids: Variations on a theme by Stokes and Einstein}\/} 2013 {\em J. Chem.
  Phys.\/} {\bf 138}, 12A526

\bibitem{Sengupta:2013dg}
Sengupta S, Karmakar S, Dasgupta C and Sastry S {\em {Breakdown of the
  Stokes-Einstein relation in two, three, and four dimensions}\/} 2013 {\em J.
  Chem. Phys.\/} {\bf 138}, 12A548

\bibitem{Kawasaki:2013bg}
Kawasaki T and Onuki A {\em {Slow relaxations and stringlike jump motions in
  fragile glass-forming liquids: Breakdown of the Stokes-Einstein relation}\/}
  2013 {\em Phys. Rev. E\/} {\bf 87}, 012312

\bibitem{Kawasaki:2014ky}
Kawasaki T, Kim K and Onuki A {\em {Dynamics in a tetrahedral network
  glassformer: Vibrations, network rearrangements, and diffusion}\/} 2014 {\em
  J. Chem. Phys.\/} {\bf 140}, 184502

\bibitem{Henritzi:2015jpa}
Henritzi P, Bormuth A, Klameth F and Vogel M {\em {A molecular dynamics
  simulations study on the relations between dynamical heterogeneity,
  structural relaxation, and self-diffusion in viscous liquids.}\/} 2015 {\em
  J. Chem. Phys.\/} {\bf 143}, 164502

\bibitem{Saw:2015ii}
Saw S and Harrowell P {\em {The geometric mean squared displacement and the
  Stokes-Einstein scaling in a supercooled liquid}\/} 2015 {\em J. Chem.
  Phys.\/} {\bf 143}, 244502

\bibitem{Ozawa:2016bk}
Ozawa M, Kim K and Miyazaki K {\em {Tuning pairwise potential can control the
  fragility of glass-forming liquids: from a tetrahedral network to isotropic
  soft sphere models}\/} 2016 {\em J. Stat. Mech.\/} {\bf 2016}, 074002

\bibitem{Schober:2016im}
Schober H~R and Peng H~L {\em {Heterogeneous diffusion, viscosity, and the
  Stokes-Einstein relation in binary liquids}\/} 2016 {\em Phys. Rev. E\/} {\bf
  93}, 052607

\bibitem{Parmar:2017hl}
Parmar A~D~S, Sengupta S and Sastry S {\em {Length-Scale Dependence of the
  Stokes-Einstein and Adam-Gibbs Relations in Model Glass Formers}\/} 2017 {\em
  Phys. Rev. Lett.\/} {\bf 119}, 056001

\bibitem{Puosi:2018cm}
Puosi F, Pasturel A, Jakse N and Leporini D {\em {Communication: Fast dynamics
  perspective on the breakdown of the Stokes-Einstein law in fragile
  glassformers}\/} 2018 {\em J. Chem. Phys.\/} {\bf 148}, 131102

\bibitem{Kammerer:1997ku}
K{\"a}mmerer S, Kob W and Schilling R {\em {Dynamics of the rotational degrees
  of freedom in a supercooled liquid of diatomic molecules}\/} 1997 {\em Phys.
  Rev. E\/} {\bf 56}, 5450--5461

\bibitem{Lombardo:2006jq}
Lombardo T~G, Debenedetti P~G and Stillinger F~H {\em {Computational probes of
  molecular motion in the Lewis-Wahnstrom model for ortho-terphenyl.}\/} 2006
  {\em J. Chem. Phys.\/} {\bf 125}, 174507

\bibitem{Chong:2009ci}
Chong S~H and Kob W {\em {Coupling and Decoupling between Translational and
  Rotational Dynamics in a Supercooled Molecular Liquid}\/} 2009 {\em Phys.
  Rev. Lett.\/} {\bf 102}, 392

\bibitem{Chen:2006kk}
Chen S~H, Mallamace F, Mou C~Y, Broccio M, Corsaro C, Faraone A and Liu L {\em
  {The violation of the Stokes-Einstein relation in supercooled water.}\/} 2006
  {\em Proc. Natl. Acad. Sci. U.S.A.\/} {\bf 103}, 12974--12978

\bibitem{Becker:2006ju}
Becker S~R, Poole P~H and Starr F~W {\em {Fractional Stokes-Einstein and
  Debye-Stokes-Einstein relations in a network-forming liquid.}\/} 2006 {\em
  Phys. Rev. Lett.\/} {\bf 97}, 055901

\bibitem{Kumar:2006hx}
Kumar P {\em {Breakdown of the Stokes--Einstein relation in supercooled
  water}\/} 2006 {\em Proc. Natl. Acad. Sci. U.S.A.\/} {\bf 103}, 12955--12956

\bibitem{Kumar:2007hl}
Kumar P, Buldyrev S~V, Becker S~R, Poole P~H, Starr F~W and Stanley H~E {\em
  {Relation between the Widom line and the breakdown of the Stokes--Einstein
  relation in supercooled water}\/} 2007 {\em Proc. Natl. Acad. Sci. U.S.A.\/}
  {\bf 104}, 9575--9579

\bibitem{Mazza:2007kr}
Mazza M~G, Giovambattista N, Stanley H~E and Starr F~W {\em {Connection of
  translational and rotational dynamical heterogeneities with the breakdown of
  the Stokes-Einstein and Stokes-Einstein-Debye relations in water.}\/} 2007
  {\em Phys. Rev. E\/} {\bf 76}, 031203

\bibitem{Xu:2009hq}
Xu L, Mallamace F, Yan Z, Starr F~W, Buldyrev S~V and Eugene~Stanley H {\em
  {Appearance of a fractional Stokes-Einstein relation in water and a
  structural interpretation of its onset}\/} 2009 {\em Nat. Phys.\/} {\bf 5},
  565--569

\bibitem{Banerjee:2009db}
Banerjee D, Bhat S~N, Bhat S~V and Leporini D {\em {ESR evidence for 2
  coexisting liquid phases in deeply supercooled bulk water}\/} 2009 {\em Proc.
  Natl. Acad. Sci. U.S.A.\/} {\bf 106}, 11448--11453

\bibitem{Mallamace:2010uj}
Mallamace F, Branca C, Corsaro C, Leone N, Spooren J, Stanley H~E and Chen S~H
  {\em {Dynamical Crossover and Breakdown of the Stokes-Einstein Relation in
  Confined Water and in Methanol-Diluted Bulk Water}\/} 2010 {\em J. Phys.
  Chem. B\/} {\bf 114}, 1870--1878

\bibitem{Jana:2011fj}
Jana B, Singh R~S and Bagchi B {\em {String-like propagation of the
  5-coordinated defect state in supercooled water: molecular origin of dynamic
  and thermodynamic anomalies}\/} 2011 {\em Phys. Chem. Chem. Phys.\/} {\bf
  13}, 16220--16226

\bibitem{Qvist:2012gg}
Qvist J, Mattea C, Sunde E~P and Halle B {\em {Rotational dynamics in
  supercooled water from nuclear spin relaxation and molecular simulations}\/}
  2012 {\em J. Chem. Phys.\/} {\bf 136}, 204505

\bibitem{Rozmanov:2012ja}
Rozmanov D and Kusalik P~G {\em {Transport coefficients of the TIP4P-2005 water
  model}\/} 2012 {\em J. Chem. Phys.\/} {\bf 136}, 044507

\bibitem{Bove:2013em}
Bove L~E, Klotz S, Str{\"a}ssle T, Koza M, Teixeira J and Saitta A~M {\em
  {Translational and Rotational Diffusion in Water in the Gigapascal Range}\/}
  2013 {\em Phys. Rev. Lett.\/} {\bf 111}, 185901

\bibitem{Dehaoui:2015ii}
Dehaoui A, Issenmann B and Caupin F {\em {Viscosity of deeply supercooled water
  and its coupling to molecular diffusion.}\/} 2015 {\em Proc. Natl. Acad. Sci.
  U.S.A.\/} {\bf 112}, 12020--12025

\bibitem{Guillaud:2017bk}
Guillaud E, Merabia S, de~Ligny D and Joly L {\em {Decoupling of viscosity and
  relaxation processes in supercooled water: a molecular dynamics study with
  the TIP4P/2005f model}\/} 2017 {\em Phys. Chem. Chem. Phys.\/} {\bf 19},
  2124--2130

\bibitem{Guillaud:2017ey}
Guillaud E, Joly L, de~Ligny D and Merabia S {\em {Assessment of elastic models
  in supercooled water: A molecular dynamics study with the TIP4P/2005f force
  field}\/} 2017 {\em J. Chem. Phys.\/} {\bf 147}, 014504

\bibitem{Galamba:2017eq}
Galamba N {\em {On the hydrogen-bond network and the non-Arrhenius transport
  properties of water.}\/} 2017 {\em J. Phys.: Condens. Matter\/} {\bf 29},
  015101

\bibitem{Kawasaki:2017gw}
Kawasaki T and Kim K {\em {Identifying time scales for violation/preservation
  of Stokes-Einstein relation in supercooled water}\/} 2017 {\em Sci. Adv.\/}
  {\bf 3}, e1700399

\bibitem{Shi:2018gu}
Shi R, Russo J and Tanaka H {\em {Origin of the emergent fragile-to-strong
  transition in supercooled water}\/} 2018 {\em Proc. Natl. Acad. Sci.
  U.S.A.\/} {\bf 115}, 9444--9449

\bibitem{MonterodeHijes:2018ec}
Montero~de Hijes P, Sanz E, Joly L, Valeriani C and Caupin F {\em {Viscosity
  and self-diffusion of supercooled and stretched water from molecular dynamics
  simulations}\/} 2018 {\em J. Chem. Phys.\/} {\bf 149}, 094503

\bibitem{Saito:2018dn}
Saito S, Bagchi B and Ohmine I {\em {Crucial role of fragmented and isolated
  defects in persistent relaxation of deeply supercooled water}\/} 2018 {\em J.
  Chem. Phys.\/} {\bf 149}, 124504

\bibitem{Kawasaki:2018vv}
Kawasaki T and Kim K {\em {Resolving spurious violation of the
  Stokes-Einstein-Debye relation in supercooled water}\/} 2018 {\em arXiv\/}
  (\textit{Preprint} \eprint{1811.00373v2})

\bibitem{Yamamoto:1997gu}
Yamamoto R and Onuki A {\em {Kinetic Heterogeneities in a Highly Supercooled
  Liquid}\/} 1997 {\em J. Phys. Soc. Jpn.\/} {\bf 66}, 2545--2548

\bibitem{Shiba:2012hm}
Shiba H, Kawasaki T and Onuki A {\em {Relationship between bond-breakage
  correlations and four-point correlations in heterogeneous glassy dynamics:
  configuration changes and vibration modes.}\/} 2012 {\em Phys. Rev. E\/} {\bf
  86}, 041504

\bibitem{Shiba:2016bi}
Shiba H, Yamada Y, Kawasaki T and Kim K {\em {Unveiling Dimensionality
  Dependence of Glassy Dynamics: 2D Infinite Fluctuation Eclipses Inherent
  Structural Relaxation}\/} 2016 {\em Phys. Rev. Lett.\/} {\bf 117}, 245701

\bibitem{Shiba:2018ip}
Shiba H, Keim P and Kawasaki T {\em {Isolating long-wavelength fluctuation from
  structural relaxation in two-dimensional glass: cage-relative
  displacement}\/} 2018 {\em J. Phys.: Condens. Matter\/} {\bf 30}, 094004

\bibitem{Plimpton:1995wl}
Plimpton S {\em {Fast parallel algorithms for short-range molecular
  dynamics}\/} 1995 {\em J. Comput. Phys.\/} {\bf 117}, 1--19

\bibitem{Abascal:2005ka}
Abascal J~L~F and Vega C {\em {A general purpose model for the condensed phases
  of water: TIP4P/2005}\/} 2005 {\em J. Chem. Phys.\/} {\bf 123}, 234505

\bibitem{Abascal:2010dw}
Abascal J~L~F and Vega C {\em {Widom line and the liquid--liquid critical point
  for the TIP4P/2005 water model}\/} 2010 {\em J. Chem. Phys.\/} {\bf 133},
  234502

\bibitem{Sumi:2013fy}
Sumi T and Sekino H {\em {Effects of hydrophobic hydration on polymer chains
  immersed in supercooled water}\/} 2013 {\em RSC Adv.\/} {\bf 3}, 12743--12750

\bibitem{Overduin:2013cu}
Overduin S~D and Patey G~N {\em {An analysis of fluctuations in supercooled
  TIP4P/2005 water}\/} 2013 {\em J. Chem. Phys.\/} {\bf 138}, 184502

\bibitem{DeMarzio:2016hl}
De~Marzio M, Camisasca G, Rovere M and Gallo P {\em {Mode coupling theory and
  fragile to strong transition in supercooled TIP4P/2005 water}\/} 2016 {\em J.
  Chem. Phys.\/} {\bf 144}, 074503

\bibitem{Hamm:2016hf}
Hamm P {\em {Markov state model of the two-state behaviour of water}\/} 2016
  {\em J. Chem. Phys.\/} {\bf 145}, 134501

\bibitem{Singh:2016bu}
Singh R~S, Biddle J~W, Debenedetti P~G and Anisimov M~A {\em {Two-state
  thermodynamics and the possibility of a liquid-liquid phase transition in
  supercooled TIP4P/2005 water}\/} 2016 {\em J. Chem. Phys.\/} {\bf 144},
  144504

\bibitem{Gonzalez:2016gr}
Gonzalez M~A, Valeriani C, Caupin F and Abascal J~L~F {\em {A comprehensive
  scenario of the thermodynamic anomalies of water using the TIP4P/2005
  model}\/} 2016 {\em J. Chem. Phys.\/} {\bf 145}, 054505

\bibitem{Handle:2018cn}
Handle P~H and Sciortino F {\em {Potential energy landscape of TIP4P/2005
  water}\/} 2018 {\em J. Chem. Phys.\/} {\bf 148}, 134505

\bibitem{Rapaport:1983ib}
Rapaport D~C {\em {Hydrogen bonds in water: Network organization and
  lifetimes}\/} 1983 {\em Mol. Phys.\/} {\bf 50}, 1151--1162

\bibitem{Saito:1995gz}
Saito S and Ohmine I {\em {Translational and orientational dynamics of a water
  cluster (H2O)108 and liquid water: Analysis of neutron scattering and
  depolarized light scattering}\/} 1995 {\em J. Chem. Phys.\/} {\bf 102}, 3566

\bibitem{Luzar:1996gw}
Luzar A and Chandler D {\em {Hydrogen-bond kinetics in liquid water}\/} 1996
  {\em Nature\/} {\bf 379}, 55--57

\bibitem{Luzar:1996gx}
Luzar A and Chandler D {\em {Effect of Environment on Hydrogen Bond Dynamics in
  Liquid Water}\/} 1996 {\em Phys. Rev. Lett.\/} {\bf 76}, 928--931

\bibitem{Luzar:2000gv}
Luzar A {\em {Resolving the hydrogen bond dynamics conundrum}\/} 2000 {\em J.
  Chem. Phys.\/} {\bf 113}, 10663

\bibitem{Kumar:2007bs}
Kumar R, Schmidt J~R and Skinner J~L {\em {Hydrogen bonding definitions and
  dynamics in liquid water}\/} 2007 {\em J. Chem. Phys.\/} {\bf 126}, 204107

\bibitem{Toninelli:2005ci}
Toninelli C, Wyart M, Berthier L, Biroli G and Bouchaud J~P {\em {Dynamical
  susceptibility of glass formers: Contrasting the predictions of theoretical
  scenarios}\/} 2005 {\em Phys. Rev. E\/} {\bf 71}, 041505

\bibitem{Rahman:1964ht}
Rahman A {\em {Correlations in the Motion of Atoms in Liquid Argon}\/} 1964
  {\em Phys. Rev.\/} {\bf 136}, A405--A411

\bibitem{Kob:1997hl}
Kob W, Donati C, Plimpton S~J, Poole P~H and Glotzer S~C {\em {Dynamical
  Heterogeneities in a Supercooled Lennard-Jones Liquid}\/} 1997 {\em Phys.
  Rev. Lett.\/} {\bf 79}, 2827--2830

\bibitem{Flenner:2005es}
Flenner E and Szamel G {\em {Relaxation in a glassy binary mixture:
  Mode-coupling-like power laws, dynamic heterogeneity, and a new non-Gaussian
  parameter}\/} 2005 {\em Phys. Rev. E\/} {\bf 72}, 011205

\end{thebibliography}
\providecommand{\newblock}{}

\end{document}